\documentclass[sigconf]{acmart}

\usepackage{amsmath,amsfonts,bm}

\def\eqref#1{equation~\ref{#1}}

\def\1{\bm{1}}

\DeclareMathAlphabet{\mathsfit}{\encodingdefault}{\sfdefault}{m}{sl}
\SetMathAlphabet{\mathsfit}{bold}{\encodingdefault}{\sfdefault}{bx}{n}

\usepackage{booktabs}
\usepackage[ruled,vlined]{algorithm2e}
\usepackage{color}
\usepackage{colortbl}
\usepackage{import}
\usepackage{paralist}
\usepackage{subcaption}
\usepackage{amsmath} 
\usepackage{caption}
\usepackage{multirow}
\usepackage{makecell}
\usepackage{threeparttable}
\usepackage{graphicx} 
\usepackage{enumitem}
\usepackage{mathtools}
\usepackage{tabularx}
\setlist[itemize]{leftmargin=*}

\AtBeginDocument{%
  \providecommand\BibTeX{{%
    \normalfont B\kern-0.5em{\scshape i\kern-0.25em b}\kern-0.8em\TeX}}}

\setcopyright{acmcopyright}
\copyrightyear{2018}
\acmYear{2018}
\acmDOI{XXXXXXX.XXXXXXX}

\acmConference[Conference acronym 'XX]{Make sure to enter the correct
  conference title from your rights confirmation emai}{June 03--05,
  2018}{Woodstock, NY}
\acmPrice{15.00}
\acmISBN{978-1-4503-XXXX-X/18/06}

\begin{document}

\title{Towards Multi-Subsession Conversational Recommendation}


\author{Yu Ji}
\authornote{Both authors contributed equally to this research.}
\affiliation{%
  \institution{Tongji University}
  \city{Shanghai}
  \country{China}}
\email{2230779@tongji.edu.cn}

\author{Qi Shen}
\authornotemark[1]
\affiliation{%
  \institution{Tongji University}
  \city{Shanghai}
  \country{China}}
\email{1653282@tongji.edu.cn}

\author{Shixuan Zhu}
\affiliation{%
  \institution{Tongji University}
  \city{Shanghai}
  \country{China}}
\email{2130768@tongji.edu.cn}

\author{Hang Yu}
\affiliation{%
  \institution{Tongji University}
  \city{Shanghai}
  \country{China}}
\email{2053881@tongji.edu.cn}

\author{Yiming Zhang}
\affiliation{%
  \institution{Tongji University}
  \city{Shanghai}
  \country{China}}
\email{2030796@tongji.edu.cn}

\author{Chuan Cui}
\affiliation{%
  \institution{Tongji University}
  \city{Shanghai}
  \country{China}}
\email{cuichuan@tongji.edu.cn}

\author{Zhihua Wei}
\authornote{Corresponding author.}
\affiliation{%
  \institution{Tongji University}
  \city{Shanghai}
  \country{China}}
\email{zhihua_wei@tongji.edu.cn}

\begin{abstract}

Conversational recommendation systems (CRS) could acquire dynamic user preferences towards desired items through multi-round interactive dialogue. 
Previous CRS  works mainly focus on the single conversation (\textbf{subsession}) that the user quits after a successful recommendation, neglecting the common scenario where the user has multiple conversations (\textbf{multi-subsession}) over a short period.
Therefore, we propose a novel conversational recommendation scenario named \textbf{M}ulti-\textbf{S}ubsession \textbf{M}ulti-round \textbf{C}onversational \textbf{R}ecommendation (\textbf{MSMCR}), where the user would still resort to CRS after several subsessions and might preserve vague interests, and the system would proactively ask attributes to activate user interests in the current subsession.
To fill the gap in this new CRS scenario, we devise a novel framework called \textbf{M}ulti-\textbf{S}ubsession \textbf{C}onversational Recommender with \textbf{A}ctivation \textbf{A}ttributes (\textbf{MSCAA}). 
Specifically, we first develop a context-aware recommendation module, comprehensively modeling user interests from historical interactions, previous subsessions, and feedback in the current subsession.
Furthermore, an attribute selection policy module is proposed to learn a flexible strategy for asking appropriate attributes to elicit user interests.
Finally, we design a conversation policy module to manage the above two modules to decide actions between asking and recommending.
Extensive experiments on four datasets verify the effectiveness of our MSCAA framework for the proposed MSMCR setting.


\end{abstract}





\maketitle
\vspace{-0.2cm}
\section{Introduction}
\vspace{-0.1cm}

Recent years have witnessed increasing attention and research effort in the conversational recommendation system (CRS), which aims to elicit dynamic user preferences and make successful recommendations through a real-time multi-round conversation with the user \cite{crssurvey}. 
To tackle this trending research topic, researchers have proposed various methods based on different conversational recommendation problem settings \cite{Abs_Greedy,QBRS,DeepCRS,EAR}. 
Among these studies, the multi-round conversational recommendation (MCR) setting is envisioned as the most realistic CRS setting so far \cite{Unicorn,MCMIPL}. 
In this work, we focus on the MCR setting where the system takes actions based on the user's current needs in each turn, either asking questions about user preferences on attributes or recommending items, with the purpose of recommending successfully within fewer turns.

Although existing MCR studies have made significant progress, they all focus on a single conversation episode that the user quits after receiving the satisfactory item recommendation \cite{EAR,SCPR,Unicorn}. 
We argue that this single conversation setting overlooks the prevalence of \textbf{multiple} conversations in the real-world CRS scenario.
The user would continue the dialogue with the system after a successful recommendation to browse items on other topics aimlessly or obtain additional system suggestions.
Under the situation of multiple conversations, the user might also preserve possible \textbf{vague} interest rather than clearly express his/her preference, which is more realistic in the world and breaks the assumption of the previous MCR works \cite{EAR,SCPR}.

\begin{figure}[t]
    \centering
    \begin{subfigure}{0.91\linewidth}
        \includegraphics[width=\textwidth]{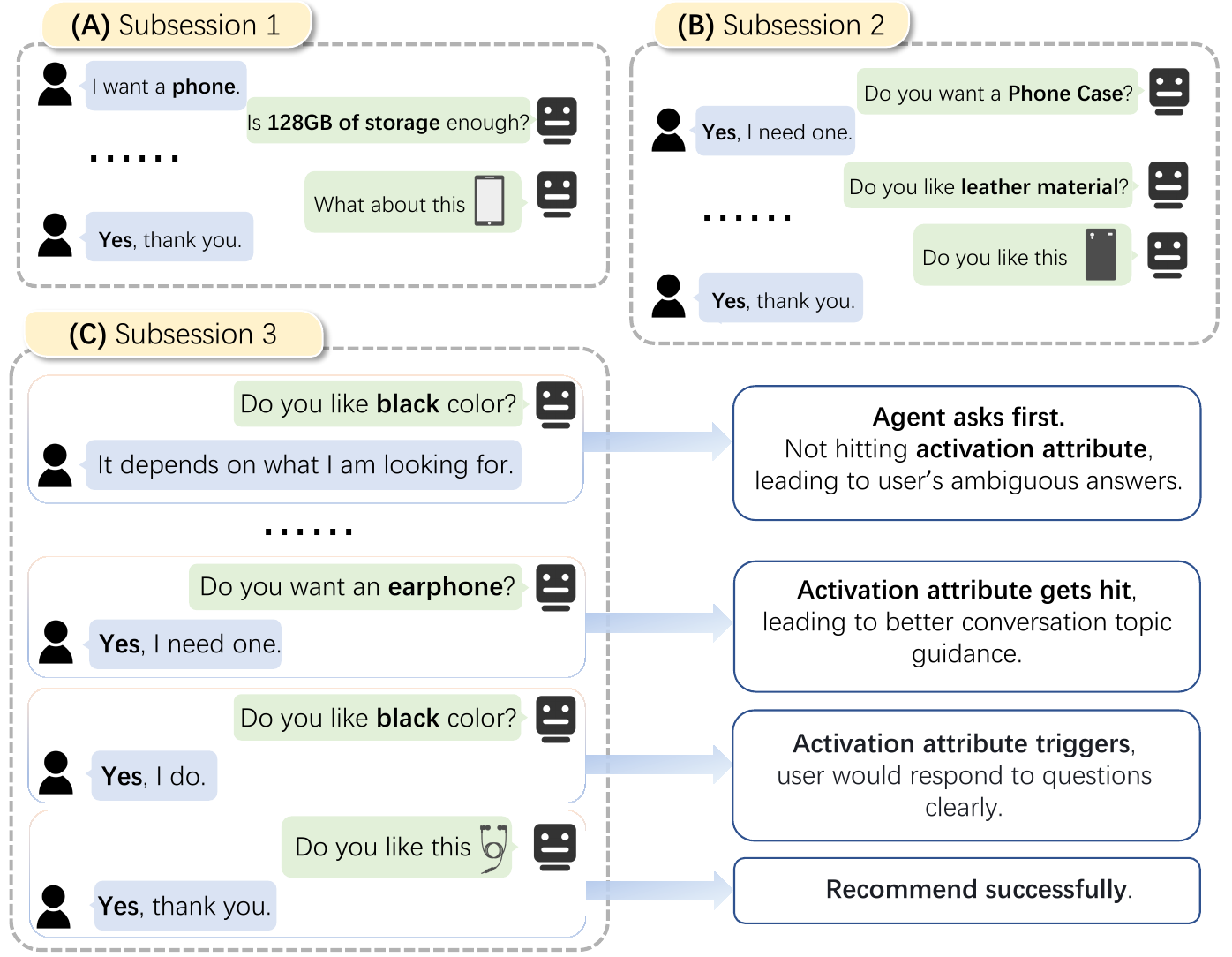}
    \end{subfigure}
    \vspace{-0.2cm}
    \captionsetup{font=small,labelfont=bf}
    \caption{A toy example of Multi-Subsession Multi-round Conversational Recommendation.
    The lower right side explains the multi-round process of the $3$rd subsession (C) of one session in detail.}
    \label{fig:toy-example}
    \vspace{-0.4cm}
\end{figure}

Figure \ref{fig:toy-example} illustrates an interaction process (\textbf{session}) that the user has three conversations (\textbf{subsessions}). 
In the first subsession of one session, the user successfully seeks his/her target item \emph{Phone}. 
At this time, he/she might require other items related to \emph{Phone} but be unaware of the related attributes to find them.
After several system suggestions, he/she further gets satisfactory items (\emph{Phone Case} and \emph{Earphone}) in the subsequent subsessions. 
More specifically considering vague user interest shown in the $3$rd subsession (C) of Figure \ref{fig:toy-example}, the user would be overwhelmed by over-specific questions (e.g., \emph{black} color) before the main attributes (e.g., \emph{earphone}) of the target item, i.e., \textbf{activation attributes} are determined, which leads to ambiguous answers (e.g., \emph{It depends on what I am looking for.} for attribute question \emph{black} color). 
Once the activation attribute \emph{earphone} is confirmed by the user as the topic of the current subsession, 
he/she would activate his/her interest and respond explicitly on system-asked questions towards the current topic, naturally including over-specific questions in the early turns.

To this end, we extend MCR \cite{EAR,SCPR,MCMIPL} to a general setting, i.e., \textbf{M}ulti-\textbf{S}ubsession \textbf{M}ulti-round \textbf{C}onversational \textbf{R}ecommendation (\textbf{MSMCR}). 
In this scenario, the user has several subsessions with the system over a short period. 
And the system would proactively ask questions on attributes to elicit dynamic user interests towards desired items in the subsequent subsessions (i.e., after the first subsession of one session).
Since our MSMCR scenario involves the user's previous subsessions and activation attributes of the current subsession, surpassing the previous MCR works, we summarize two main challenges to handle this scenario listed as follows:
\begin{itemize}[leftmargin=*]
    \item \textbf{How to model user interests comprehensively?} 
    Except for historical user-item interactions and feedback in the current subsession, the additional previous subsessions information should be further considered in our MSMCR scenario.
    As such, how to organize and aggregate these multiple information sources effectively for modeling user interests is a problem.
    \item \textbf{How to ask appropriate questions?} In our scenario, the system is required to ask appropriate questions to elicit the user’s interest for recommending his/her desired item. 
    Concretely, the user preserves unclear interest at the beginning of the current subsession, where the system should activate his/her current interest as early as possible. 
    After that, the system needs to ask attributes to eliminate the uncertainty of candidate items, following existing MCR works \cite{EAR,SCPR}. 
    As a result, it is essential to learn a flexible attribute selection strategy to activate user interests and then reduce the uncertainty of candidate items. 
\end{itemize}

To effectively address the aforementioned challenges for MSMCR, we propose a novel method named \textbf{M}ulti-\textbf{S}ubsession \textbf{C}onversational Recommender with \textbf{A}ctivation \textbf{A}ttributes (\textbf{MSCAA}), which organically combines the two policy modules with a recommendation module.  
In detail, \textit{Context-aware Recommendation} module models multiple information for item and attribute prediction. 
More specifically, it extracts user long-term interest representations from the user-item-attribute graph via a relational graph convolution network.
Then for previous online subsessions information, we adopt dual gated recurrent units to learn user short-term interest representations. 
Besides, the user feedback, i.e., accepted attributes, rejected attributes and items in the current subsession, is also aggregated to represent the user's current interest.  
Finally, a unified transformer encoder is applied to fuse the above three different interests for prediction.
The \textit{Attribute Selection Policy} module automatically learns the strategy for asking activation attributes or others based on predicted user-like score and entropy of candidate attributes.
And enventually, the \textit{Conversation Policy} module further utilizes the above two modules and then decides actions between asking and recommending.

Our main contributions of this work are summarized below:
\begin{itemize}[leftmargin=*]
    \item We introduce a novel multi-subsession conversation recommendation scenario named MSMCR, a natural extension of the multi-round conversational recommendation, which considers previous online subsessions and system-guided user interest activation.
    \item Building upon our new scenario, we devise a novel framework MSCAA that combines the user's long- and short-term interaction and real-time feedback to model user interests comprehensively and learn a flexible asking policy to activate and crystallize user demands.
    \item We evaluate MSCAA on four adapted datasets and conduct a human evaluation to demonstrate the effectiveness of our method.
\end{itemize}

\vspace{-0.2cm}
\section{Related Works} \label{sec:relate_works}
\vspace{-0.1cm}

\noindent \textbf{Conversational Recommendation.}
Conversational recommendation systems (CRS) \cite{crssurvey,crstu} leverage immediate user feedback information to capture dynamic user preferences and make precise recommendations in conversational scenarios. 
Existing CRS research can be categorized into four directions.
(1) Multi-round Conversational Recommendation (MCR) \cite{EAR,CRM}. In each dialogue turn, MCR chooses between attribute asking and item recommendation to achieve an early and accurate hit on the target item. 
(2) Exploration-Exploitation Trade-offs \cite{li2021seamlessly,zhao2022knowledge}. These methods focus on the cold-start situation and propose a paradigm to balance users’ exploration and exploitation trade-offs. 
(3) Question-based User Preference Elicitation works \cite{SAUR,QBRS} ask ``clarification/clarifying'' questions to figure out how to gain more preference information within limited conversations. 
(4) Dialogue Understanding and Generation \cite{DeepCRS,zhou2020Topic} focuses on how to understand and provide a real conversation and make a smoother response.

\noindent \textbf{Multi-round Conversational Recommendation.} 
Among all the above CRS research scenarios, we focus on the MCR task, in which the system alternates between asking attribute-based questions and recommending items several times to seek the user's desired item within fewer interaction turns.
The conversation strategy is the core design in MCR, for deciding whether to ask or recommend, and the strategy is typically modeled by deep reinforcement learning (DRL) \cite{PG,DQN} to tackle this multi-step decision-making process.
For instance, CRM \cite{CRM} brings up the conversation setting starting with the agent's query, which gathers user feedback information and gives recommendations in the end.
EAR \cite{EAR} extends the CRM framework where the agent can recommend many times.
SCPR \cite{SCPR} proposes an interactive path reasoning method to find appropriate candidates for items and attributes.
FPAN \cite{FPAN} utilizes online user feedback to update user embedding via the gate mechanism based on EAR framework.
UNICORN \cite{Unicorn} contributes a dynamic weighted graph-based unified policy learning framework.
Furthermore, CRIF \cite{CRIF} explicitly uses item feedback to capture the implicit user preference for the recommendation and employs inverse reinforcement learning to learn a better conversation strategy.
Considering RL often suffers from unstable learning, some recent works attempt non-RL alternative methods, such as a simple rule \cite{Minicorn} or decision tree \cite{FacTCRS},  to achieve comparable performance.
In addition to the above classical MCR task, more and more studies start to focus on exploring the new MCR scenario applied in many fields.
For example, MIMCR \cite{MCMIPL} considers the incompleteness and diversity of user interests and extends MCR to a more realistic setting named multiple choice questions, which is adopted by follow-ups \cite{zhao2023towards,kim2023confident}. 
Bundle-MCR \cite{BundleMCR} incorporates bundle generation with MCR mechanisms.
GPR \cite{GPR} proposes a two-level graph path reasoning method that also recommends items when asking attributes.
MetaCRS \cite{MetaCRS} contributes a meta reinforcement learning framework for the scenario of cold-start users.
Different from these works, and some studies \cite{PreCRS,SeqCR} that utilize the user's historical interaction sequence to improve recommendations in MCR, we concentrate on a \textbf{novel scenario} where the user has several dialogues with the system in a short time and might preserve unclear interests.
\vspace{-0.1cm}
\section{MSMCR scenario}\label{sec:define}
\vspace{-0.05cm}

\subsection{Definition}
\vspace{-0.05cm}

In this scenario, we define the set of users $\mathcal{U}$ and items $\mathcal{V}$. We collect all attributes $\mathcal{A}$ corresponding to items, and each item $v$ is associated with a set of attributes $\mathcal{A}_v$. 
Unlike previous MCR works that call a conversation session, we define a conversation episode as a \emph{subsession} $s$. 
And a \emph{session} $S_u = [s^1_u, s^2_u, \ldots, s^{n-1}_u, s^n_u]$ consists of multiple subsessions where $n$ is variational length for each session.
Moreover, a session can be divided into two parts: the previous subsessions $ P_{S_u} = [s^1_u, s^2_u, \ldots, s^{n-1}_u]$ and the current subsession $s^n_u$. 
The goal of MSMCR is to recommend the desired item $v^n$ of user $u$ in the current subsession $s^n_u$ within fewer turns, based on $P_{S_u}$ and the current subsession information.

For a user $u \in \mathcal{U}$, the workflow of MSMCR is listed as follows: 
\textbf{(1)} After several subsessions (i.e., $P_{S_u}$) successfully end, the user resorts to the subsequent conversation (i.e., $s_u^n$) instead of quitting the session. In this state, the user might initially have no explicit attribute query, and hence the current subsession is started from the system side.
\textbf{(2)} Then, the system is free to \emph{ask} questions about an attribute from the candidate attribute set $\mathcal{A}_{cand}$ (e.g., $\mathcal{A}_{cand}=\mathcal{A}$ at first) or to \emph{recommend} a certain number of items (e.g., top-$K$) from the candidate item set $\mathcal{V}_{cand}$.
\textbf{(3)} Next, user $u$ provides feedback, i.e., accept, \textbf{unknown}, or reject for the asked attribute, and accept or reject for the recommended items.
\textbf{(4)} After that, the system updates the sets of candidate attributes and items based on user feedback.
\textbf{(5)} Within multiple iterations of step \textbf{(2)}-\textbf{(4)}, the system elicits clearer user interests and provides more accurate recommendations. 
The current subsession will terminate when the system recommends successfully, or the interaction turn reaches the maximum $T$.

Different from the user setting that has clear attribute preferences in previous MCR works \cite{EAR,SCPR}, the user's preference might be vague and should be proactively guided \cite{zhou2020Topic} by the system in the MSMCR scenario.
That is, in step \textbf{(3)}, the user will respond \textbf{``unknown''} for the other attributes when the \textbf{activation attributes} $\mathcal{A}_{v^n}^*$ (i.e., several main attributes that enable the user to trigger his/her current demands) have not been clarified.
And after one of the activation attributes is hit by the system, the user would generate explicit feedback (accept or reject) for all attributes.

\begin{figure*}
\vspace{-0.3cm}
    \centering
    \includegraphics[width=0.8\textwidth]{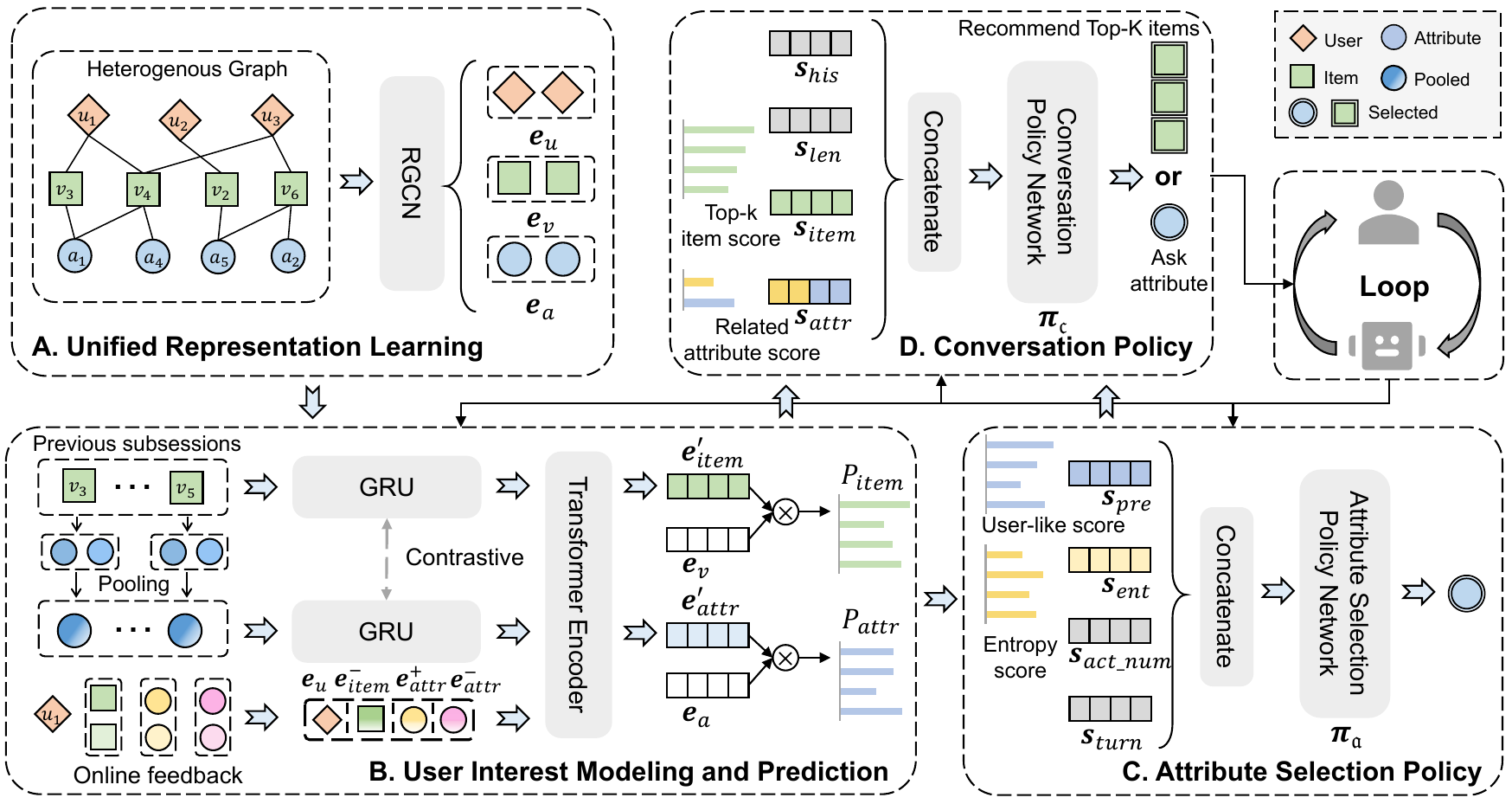}
    \vspace{-0.2cm}
    \caption{The overview of MSCAA. 
        Our framework includes recommendation modules (A and B) and policy modules (C and D). }
    \label{fig:framework}
    \vspace{-0.2cm}
\end{figure*}

\vspace{-0.15cm}
\subsection{General Framework} \label{gen-frame}
\vspace{-0.15cm}
Similar to existing MCR works, we formulate our MSMCR problem as a Markov Decision Process (MDP) of interaction between the user and the recommendation system agent. 
The goal of our framework is to learn the policy network $\pi_{\mathfrak c}$ (cf. Section \ref{sec:agent-conv}) to maximize the expected cumulative rewards for the overall conversations.
We decompose one conversation turn into four steps: \textbf{state}, \textbf{action}, \textbf{transition} and \textbf{reward} under our framework as follows.

\noindent \textbf{State.} The system maintains the conversation state ${s}_t$ at each turn $t$ in the subsession, which encodes the user feedback and candidate item information. Specially, user $u$'s feedback includes the accepted attributes $\mathcal{A}_{acc}^t$, rejected attributes $\mathcal{A}_{rej}^t$ and rejected items $\mathcal{V}_{rej}^t$. 

\noindent \textbf{Action.} 
The agent takes actions according to the conversation state, where the action can be asking an attribute or recommending items. 
If the system decides to ask, it will select an attribute from the decision of an attribute agent using policy $\pi_{\mathfrak a}$ (cf. Section  \ref{sec:agent-attr}). 
While for recommendation action, the system will select top-$K$ items based on the recommendation score (cf. Section \ref{sec:rec}). 

\noindent \textbf{Transition.}
When the agent takes actions, and the user provides corresponding feedback to the attribute or items, the current state will transition to the next $s_{t+1}$.
In this step, we update the sets $\mathcal{A}_{acc}$, $\mathcal{A}_{rej}$, $\mathcal{V}_{rej}$, $\mathcal{V}_{cand}$ and $\mathcal{A}_{cand}$ based on user feedback following the transition operation of \cite{MCMIPL,HICR}. 
Note that in our MSMCR setting, user feedback for the asked attribute will be ``unknown'' until an activation attribute is hit. In this case, we do not update them. 

\noindent \textbf{Reward.} 
We design six kinds of rewards in our framework: (1) $r_{rec\_acc}$ and (2) $r_{rec\_rej}$ denote positive and negative rewards when the user accepts or rejects the recommended item, respectively; (3) $r_{ask\_acc}$, (4) $r_{ask\_rej}$ and (5) $r_{ask\_unk}$ denote positive, negative and negative rewards when the user accepts, rejects and responds ``unknown'' to  the asked attribute, respectively; (6) $r_{quit}$ is a strongly negative reward if the maximum subsession turn $T$ is reached.

\vspace{-0.15cm}
\section{Methodology}\label{sec:method}
\vspace{-0.15cm}
As shown in Figure \ref{fig:framework}, our framework contains three main components:
(1) \textit{Context-aware Recommendation} module including Unified Representation Learning (A) and User Interest Modeling and Prediction (B), which comprehensively models user interests based on the information from historical interactions, previous subsessions, and the current subsession;
(2) \textit{Attribute Selection Policy} module (C), which aims to select an appropriate asked attribute based on attribute-related information;
(3) \textit{Conversation Policy} module (D), which decides actions between asking and recommending based on the overall conversation.

\vspace{-0.2cm}
\subsection{Context-aware Recommendation}\label{sec:rec}
\vspace{-0.02cm}
A well-learned recommendation module will improve the performance of the online system agent \cite{CRIF}. To this end, we carefully design the following three parts in the context-aware recommendation: \emph{unified representation learning} to extract long-term user interest, \emph{user interest modeling} to learn short-term interest from subsessions (i.e., context) and then integrate different interests, and finally, \emph{prediction} to estimate probabilities that how the user likes items and attributes.

\vspace{-0.2cm}
\subsubsection{Unified Representation Learning.}\label{sec:ui-encoder}
We first construct a unified heterogeneous graph \cite{MCMIPL,pang} based on historical user-item interactions and static attribute-item relations, including three kinds of nodes: users, items, and attributes. This graph can be denoted as $G = \{N = \{\mathcal{U}, \mathcal{V}, \mathcal{A}\}, E = \{\mathcal{E}_{uv}, \mathcal{E}_{av}\}\}$, where the user-item edge $(u, v) \in \mathcal{E}_{uv}$ denotes that user $u$ has interacted with item $v$ and the attribute-item edge $(a, v) \in \mathcal{E}_{av}$ means that attribute $a$ belongs to item $v$.
Then, we introduce a $L_g$-layer relational graph convolutional network (RGCN) \cite{rgcn} to learn the node representations in the graph $G$. At first, each user, item, and attribute node is assigned with a unique node index and then initialized with the node embedding matrix $\boldsymbol{E}^0 \in \mathbb{R}^{\left|N\right| \times D}$, as the input embeddings of the first layer of RGCN.
And for each layer until $L_g$, 
RGCN will first model different edge types based on node neighbors, and then the aggregated edge features will be integrated into node features. We take item node $v$ which involves two kinds of edges $\mathcal{R}_v = \{ r_{uv}, r_{av} \}$ as an example as follows: 
\begin{equation}
\setlength{\abovedisplayskip}{2pt}
\setlength{\belowdisplayskip}{2pt}
\begin{aligned}
\boldsymbol{e}_{v}^{l+1}={\rm ReLU}(\sum\nolimits_{r \in \mathcal{R}_v} \sum\nolimits_{i \in \mathcal{N}_{v}^{r}} \frac{1}{\sqrt{\left| \mathcal{N}^r_v \right| \left| \mathcal{N}_i^r\right|}} \boldsymbol{W}_{r}^{l} \boldsymbol{e}_{i}^{l}+\boldsymbol{W}_{0}^{l} \boldsymbol{e}_{v}^{l}),
\end{aligned}\label{RGCN}
\end{equation}
where $\mathcal{N}_{v}^{r}$ denotes the neighbor nodes of node $v$ under relation $r$. 
$\boldsymbol{W}_{0}^{l}$, $\boldsymbol{W}_{r_{uv}}^{l}$ and $\boldsymbol{W}_{r_{av}}^{l} \in \mathbb{R}^{D \times D}$ are trainable parameters.
We can obtain the refined user and attribute embeddings $\boldsymbol{e}_{u}^{l+1}$, $\boldsymbol{e}_{a}^{l+1}$ in the same way. 
After $L_g$ layers information propagation, we combine node embeddings of each layer to capture different semantics \cite{he2020lightgcn}: 
\begin{equation}
\setlength{\abovedisplayskip}{2pt}
\setlength{\belowdisplayskip}{2pt}
\begin{aligned}
\boldsymbol{e}_{n}=\frac{1}{L_{g}+1} \sum\nolimits_{j=0}^{L_g} \boldsymbol{e}_{n}^{j}, \;\;\forall n \in N = \{\mathcal{U}, \mathcal{V}, \mathcal{A}\}.
\end{aligned}\label{RGCN-agg}
\end{equation}

In the rest of this paper, we will use $\boldsymbol{e}_{u}$, $\boldsymbol{e}_{v}$ and $\boldsymbol{e}_{a}$ to represent refined user, item and attribute embedding, respectively. 
Note that $\boldsymbol{e}_{u}$ can be regarded as \textbf{long-term} user interest \cite{FPAN,pang}.

\vspace{-0.2cm}
\subsubsection{User Interest Modeling.}\label{sec:interest-model}

In our scenario, 
the user also has previous subsessions that occur recently with the system online, which are probably relevant, and hence crucial for user short-term interest modeling.
For each subsession $s^i_u$ in $P_{S_u}$, much abundant information should be considered, including the accepted item $v^i$ and its associated attributes $\mathcal{A}_{v^i}$, which reflects short-term user preference on item and attributes.
Therefore, we have two sequences from $P_{S_u}$: the previous desired item sequence $P^{\mathcal{V}}_{S_u} = [v^1, v^2, \ldots, v^{n-1}]$ and a sequence of the previous desired attribute sets $P^{\mathcal{A}}_{S_u} = [\mathcal{A}_{v^1}, \mathcal{A}_{v^2}, \ldots, \mathcal{A}_{v^{n-1}}]$. 
Following the classic sequential recommendation method \cite{GRU4Rec,DREAM},
we separately model the above two sequences via gated recurrent units (GRU) \cite{GRU} to capture the temporal interactions between each subsession.
Specifically, for item sequence $P^{\mathcal{V}}_{S_u}$, we use the final hidden state of the GRU as the representation of the user's \textbf{short-term item-level} interest: 
\begin{equation}
\setlength{\abovedisplayskip}{2pt}
\setlength{\belowdisplayskip}{2pt}
\begin{aligned}
\boldsymbol{e}'_{item}=\mathrm{GRU}(\{\boldsymbol{e}_v^1,\boldsymbol{e}_v^2, \dots, \boldsymbol{e}_v^{n-1}\}),
\end{aligned}\label{GRU}
\end{equation}
where $\boldsymbol{e}_v^i$ is the  representation of $i$-th item in $P^{\mathcal{V}}_{S_u}$.
Similarly, we can get the user's \textbf{short-term attributes-level} interest $\boldsymbol{e}'_{attr}$ via another GRU.
Note that $P^{\mathcal{A}}_{S_u}$ is a attribute set sequence.
Therefore, we first generate the overall representation of each attribute set by mean pooling operation:
$\boldsymbol{g}_{a}^{i} = {\rm {MEAN}}(\{ \boldsymbol{e}_{a} | a \in \mathcal{A}_{v^i}  \})$  \cite{DREAM}.

For the current subsession, the online feedback including the accepted and rejected attributes and the rejected items, is also pivot information for revealing the \textbf{current} user interest.
To comprehensively integrate the user's long- and short-term interest as well as current interest, we adopt a unified $L_t$-layer Transformer encoder \cite{transformer} to fuse these multiple kinds of information.
In our case, the input sequence is $\left[\boldsymbol{e}_u,\boldsymbol{e}'_{item},\boldsymbol{e}'_{attr},\boldsymbol{e}_{item}^{-},\boldsymbol{e}_{attr}^{+},\boldsymbol{e}_{attr}^{-}\right]$, where $\boldsymbol{e}_{item}^{-},\boldsymbol{e}_{attr}^{+},\boldsymbol{e}_{attr}^{-}\in \mathbb{R}^{1 \times D}$ denotes the aggregated representation of rejected items, accepted and rejected attributes. 
Here, we use the mean pooling operation to generate these three kinds of information, e.g., 
$\boldsymbol{e}_{item}^{-} = {\rm MEAN}(\{\boldsymbol{e}_v|v\in\mathcal{V}_{rej}\})$.
For the first few turns that may have no rejected items, accepted attributes or rejected attributes, the corresponding $\boldsymbol{e}_{item}^{-},\boldsymbol{e}_{attr}^{+},\boldsymbol{e}_{attr}^{-}$  is masked for Transformer input.
Finally, we can obtain the user's \textbf{final item-level and attribute-level interest} from different kinds of information, i.e., $\boldsymbol{e}'_{item} $ and $\boldsymbol{e}'_{attr}$ in the output embeddings of Transformer encoder.
Here we maintain the notations of transformed output embeddings unchanged for the sake of simplicity.

\vspace{-0.2cm}
\subsubsection{Prediction}\label{sec:score-model}

Based on the learned representations above, we can estimate the probability that the user likes items and attributes:
\begin{equation}
\setlength{\abovedisplayskip}{2pt}
\setlength{\belowdisplayskip}{2pt}
\begin{aligned}
P_{item}(u, v)=\tanh ({\boldsymbol{e}'_{item}}^\top \boldsymbol{e}_v) \;;\; 
P_{attr}(u, a)=\tanh ({\boldsymbol{e}'_{attr}}^\top \boldsymbol{e}_a),
\end{aligned}\label{eqn:score}
\end{equation}
where $v$ and $a$ are candidate item in $\mathcal{V}_{cand}$ and attribute in $\mathcal{A}_{cand}$, respectively.

\vspace{-0.2cm}
\subsection{Attribute Selection Policy}\label{sec:agent-attr}
\vspace{-0.02cm}
In our MSMCR scenario, due to the existence of special activation attributes, we expect these activation attributes that the user might well like should be hit as soon as possible.
After that, other target attributes asked can better eliminate the uncertainty of candidate items.
However, the previous attribute selection rules, like \emph{Max Entropy} \cite{SCPR,CRIF} and \emph{Max User-like} \cite{MCMIPL}, only consider one aspect of the above issues, e.g., \emph{Max User-like} is mainly suitable for asking attributes that match user interests. 
Furthermore, these rules cannot deal with the dynamic attribute response setting well in MSMCR, especially the more complex environment in the real-world scenario.
Hence we employ an attribute selection policy to adaptively select the appropriate asked attribute to handle the above challenge. 

Specifically, the policy network is implemented by a two-layer MLP, which can output the asked attribute based on the user's conversation state $\boldsymbol{s}_{\mathfrak a}$. 
The state vector $\boldsymbol{s}_{\mathfrak a}$ involves $4$ parts of information as follows:
\begin{equation}
\setlength{\abovedisplayskip}{2pt}
\setlength{\belowdisplayskip}{2pt}
\begin{aligned}
    \boldsymbol{s}_{\mathfrak a}= \boldsymbol{s}_{pre} \oplus \boldsymbol{s}_{ent} \oplus \boldsymbol{s}_{act\_num} \oplus \boldsymbol{s}_{turn},
\end{aligned}\label{eqn:state_attr}
\end{equation}
where $\oplus$ is the concatenation operation, $\boldsymbol{s}_{pre}$ encodes user interests on all attributes, in which each dimension is the estimated probability (i.e., preference score) $P_{attr}$ calculated by Eq. (\ref{eqn:score}).
$\boldsymbol{s}_{ent}$ is the entropy of each attribute among candidate items $\mathcal{V}_{cand}$, where each dimension is the entropy $w_a$ of attribute $a$ calculated by $w_{a}=-{\rm p}(a)\log{\rm p}(a)-(1-{\rm p}(a))\log(1-{\rm p}(a))$, where ${\rm p}(a)=\left |\mathcal{V}_{cand} \cap \mathcal{V}_{a}\right | / \left |\mathcal{V}_{cand}\right |$, following \cite{EAR,CRIF}.
Note that only the attributes in $\mathcal{A}_{cand}$ are preserved for the attribute selection, and all other attributes are masked in $\boldsymbol{s}_{pre}$ and $\boldsymbol{s}_{ent}$.
These two vectors provide useful user-side and item-side information for attribute selection, following \cite{EAR}.
$\boldsymbol{s}_{act\_num}$ and $\boldsymbol{s}_{turn}$ denote the accepted attributes number and the current turn number, to perceive the changes during the conversation.
This agent receives the ask-related rewards ($r_{ask\_acc}$, $r_{ask\_rej}$ and $r_{ask\_unk}$) and user-quit reward ($r_{quit}$) only, and updates the state with corresponding state transitions.

\vspace{-0.2cm}
\subsection{Conversation Policy}\label{sec:agent-conv}
\vspace{-0.02cm}
The conversation policy is responsible for deciding whether to ask an attribute or recommend items with the purpose of hitting the user's desired item of the current subsession in fewer turns. 
We use a simple two-layer MLP to implement the interaction policy network, which maps the state vector $\boldsymbol{s}_{\mathfrak c}$ to two actions, i.e., ask or recommend. 
The action with a higher expected reward will be selected by agent. 
The state vector $\boldsymbol{s}_{\mathfrak c}$ is the concatenation of the following components:
\begin{equation}
\setlength{\abovedisplayskip}{2pt}
\setlength{\belowdisplayskip}{2pt}
\begin{aligned}
    \boldsymbol{s}_{\mathfrak c}= \boldsymbol{s}_{his} \oplus \boldsymbol{s}_{len} \oplus \boldsymbol{s}_{item} \oplus \boldsymbol{s}_{attr},
\end{aligned}\label{eqn:state_conv}
\end{equation}
where $\boldsymbol{s}_{his}$ encodes the conversation history in the subsession and $\boldsymbol{s}_{len}$ encodes the size of the candidate item set following previous CRS works \cite{EAR,SCPR}. 
$\boldsymbol{s}_{item}$ is a $K$-dimension vector which records the top-$K$ candidate item score from the context-aware recommendation module. 
$\boldsymbol{s}_{attr}$ represents the preference and entropy score of the attribute by the attribute selection policy. 
Intuitively, the greater value is in vector $\boldsymbol{s}_{item}$, the more probably the agent selects the ``recommend'' action. 
The same is for $\boldsymbol{s}_{attr}$.
By comparing the last two vectors, the agent can make full use of detailed information to decide the appropriate action.

\vspace{-0.2cm}
\subsection{Model Training}\label{sec:training}
\vspace{-0.02cm}
\subsubsection{Recommendation Pre-Training.}\label{rec-train}
Since the performance of the online agent largely depends on the context-aware recommendation module, we first pre-train this module based on training interaction data for item and attribute prediction.

For item prediction, given a session $S_u = \{P_{S_u}, s^n_u\}$, 
the target item $v^n$ is considered as a positive ground-truth item, which is expected to rank higher than other negative candidate items. Following \cite{EAR}, We employ pairwise Bayesian Personalized Ranking (BPR) \cite{bpr} loss:
\begin{equation}
\setlength{\abovedisplayskip}{2pt}
\setlength{\belowdisplayskip}{2pt}
\begin{aligned}
\mathcal{L}_{item} =\sum\nolimits_{\left(u, v^n, v^{\prime}\right) \in \mathcal{D}_{1}} -\ln \sigma\left(P_{item}\left(u,v^n \right)-P_{item}\left( u, v^\prime \right)\right),
\end{aligned}\label{eqn:item-pred-loss}
\end{equation}
where $\mathcal{D}_{1}\coloneqq\left\{\left(u, v^{n}, v^{\prime }\right) \mid v^{\prime } \in \mathcal{V} \setminus \mathcal{V}_{u}\right\}$ denotes the training set of item pairs where $\mathcal{V}_u$ is the set of items interacted by user $u$ and $v^\prime$ is sampled from non-interacted items of user $u$.
Similar to item prediction, attribute prediction needs to rank the attributes of target item $a \in \mathcal{A}_{v^n}$ ahead of others. 
To achieve this, we also adopt BPR loss as $\mathcal{L}_{attr1}$. 
Besides, it is noted that in our MSMCR scenario, activation attributes are more vital than any other attributes for triggering user interests, which should be ranked higher. 
Therefore, the loss w.r.t. activation attribute prediction is defined as:
\begin{equation}
\setlength{\abovedisplayskip}{2pt}
\setlength{\belowdisplayskip}{2pt}
\begin{aligned}
\mathcal{L}_{attr2} =\sum\nolimits_{(u, a, a^{\prime}) \in \mathcal{D}_{2}} -\ln \sigma(P_{attr}(u, a )-P_{attr}( u, a^\prime)),
\end{aligned}\label{eqn:act-attr-pred-loss}
\end{equation}
where $\mathcal{D}_{2}\coloneqq\{(u, a, a^{\prime }) \mid a \in \mathcal{A}_{v^{n}}^*, a^{\prime } \in \mathcal{A}_{v^n} \setminus \mathcal{A}_{v^n}^*\}$ denotes the pairwise attribute training data. The final loss of attribute prediction is: $\mathcal{L}_{attr} = \mathcal{L}_{attr1} + \mathcal{L}_{attr2}$.

Meanwhile, we additionally introduce a contrastive loss (InfoNCE \cite{InfoNCE}) to constrain the uniformity of user item-level and attribute-level interest:
\begin{equation}
\setlength{\abovedisplayskip}{2pt}
\setlength{\belowdisplayskip}{2pt}
\begin{aligned}
\mathcal{L}_{cont} = \sum\nolimits_{k} {-\log \frac{{\exp ({\boldsymbol{e}_{item,k}^\prime}^\top  \boldsymbol{e}_{attr, k}^\prime  / \tau)}} 
{\sum\nolimits _{k^{\prime}}\exp ({\boldsymbol{e}_{item,k}^\prime}^\top  \boldsymbol{e}_{attr, k^\prime }^\prime / \tau) }},
\end{aligned}\label{eqn:cont-loss}
\end{equation}
where $\tau$ is a temperature hyper-parameter to control the concentration of features.
For $k$-th item-level interest representation $\boldsymbol{e}_{item,k}^\prime$ within the same minibatch, $\boldsymbol{e}_{attr, k}^\prime$ is a positive sample that reflects the same user's attribute-level interest, while all others are negative samples.
By this constraint term, user interests from different views can complement each other to enhance the final interest representation.
Finally, we optimize the recommendation model with the above losses jointly: 
\begin{equation}
\setlength{\abovedisplayskip}{2pt}
\setlength{\belowdisplayskip}{2pt}
\begin{aligned}
\mathcal{L}_{rec}=\mathcal{L}_{item}+\mathcal{L}_{attr}+\omega\mathcal{L}_{cont},
\end{aligned}\label{eqn:loss}
\end{equation}
where $\omega$ controls the weight of constraint loss.

\vspace{-0.2cm}
\subsubsection{Policy Pre-Training.}
Then we pre-train the above two policies $\pi_{\mathfrak a}$ and $\pi_{\mathfrak c}$ to accelerate training by using DAgger \cite{dagger} with expert demonstrations. 
It iteratively trains policy via supervised learning on observation-action pairs $(s^*,a^*)$ from the online expert rule:
$\hat{\pi}_{i}=\arg\; {\min}_{\pi_{i}} \sum -\ln \pi_{i}(a^*|s^*)$, where $i\in \{\mathfrak a,\mathfrak c\}$.
For attribute selection policy $\pi_{\mathfrak a}$, the expert online rule is defined that if the number of accepted attributes is less than $1$, the attribute agent asks the attribute with \emph{Max User-like} otherwise with \emph{Max Entropy}.
And for conversation policy $\pi_{\mathfrak c}$, the expert rule is defined that at turn $t$, the conversation agent chooses the \emph{ask} action with probability $1 - \frac{t}{T}$ or \emph{recommends} top-$K$ items. 
The intuition behind this is that attributes are expected to be asked in the first few turns, and later the recommendation is presented.

\vspace{-0.2cm}
\subsubsection{Policy Training.}\label{agent-train}
After the recommendation and policy are pre-trained, we further train two policy modules with online user interaction meanwhile the recommendation module is fixed. 
Concretely, we utilize each session collected online to optimize these two policy networks by Policy Gradient \cite{PG}  following \cite{FPAN,BundleMCR}.

\vspace{-0.1cm}
\section{Experiments}
\vspace{-0.05cm}

In this section, we conduct experiments on the MSMCR scenario to evaluate the performance of our method compared with other adapted state-of-the-art (SOTA) models\footnote{Our code and data will be released for research purposes.}.

\newcommand{\tabincell}[2]{\begin{tabular}{@{}#1@{}}#2\end{tabular}}  
\begin{table}[htbp]
\vspace{-0.05cm}
\setlength{\abovecaptionskip}{0cm}
\setlength{\belowcaptionskip}{-0.15cm}
    \caption{Statistics of datasets used in experiments, where $n$ is the length of one session. }
    \label{tab:dataset}
    \centering
    \small
    \resizebox{0.98\columnwidth}{!}{
    \begin{tabular}
    {p{1.5cm}<{\raggedright}p{0.9cm}<{\raggedleft}p{0.8cm}<{\raggedleft}p{1.2cm}<{\raggedleft}p{1.2cm}<{\raggedleft}|
    p{1.0cm}<{\raggedleft}}
    \toprule
    Statistic& \#Users & \#Items & \#Interactions & \#Attributes &Avg. $n$   \\
    \midrule
    LastFM/LastFM*  & 1,801 & 7,432 &76,693  & 33/8438 & 2.82/2.82  \\
    Yelp/Yelp*  &27,675 & 70,311 &1,368,606 & 29/590  & 2.85/2.85  \\
    \bottomrule
    \end{tabular}}
    \vspace{-0.3cm}
\end{table}

\begin{table*}[t]
    \centering
    \setlength{\abovecaptionskip}{0cm}
\setlength{\belowcaptionskip}{-0.1cm}
    \caption{
    Experimental results on four datasets. The {\bfseries bold} number indicates the improvements over the best baseline (\underline{underlined}) are statistically significant ($p \textless 0.01$) with paired $t$-test.
    SR, hN and AR stand for SR@$\emph{10}$, hN@($\emph{10}$, $\emph{10}$) and AR@$\emph{10}$, respectively.
    All the baselines with $\dagger$ and $\dagger-$ superscript are adapted to our MSMCR setting (cf. Section \ref{sec:baseline} for more details).
    }
    \label{tab:overall}
    \resizebox{\textwidth}{!}{
    \begin{tabular}{p{1.4cm}<{\raggedright}p{0.73cm}<{\centering}p{0.73cm}<{\centering}p{0.83cm}<{\centering}p{0.82cm}<{\centering}p{0.01cm}p{0.73cm}<{\centering}p{0.73cm}<{\centering}p{0.83cm}<{\centering}p{0.82cm}<{\centering}p{0.01cm}p{0.73cm}<{\centering}p{0.73cm}<{\centering}p{0.83cm}<{\centering}p{0.82cm}<{\centering}p{0.01cm}p{0.73cm}<{\centering}p{0.73cm}<{\centering}p{0.83cm}<{\centering}p{0.82cm}<{\centering}}
    \toprule
    \multirow{2}{*}{ \bfseries Models}& \multicolumn{4}{c}{ \bfseries LastFM }& & \multicolumn{4}{c}{\bfseries LastFM* }& &\multicolumn{4}{c}{\bfseries Yelp}& \multicolumn{5}{c}{ \bfseries Yelp*}  \\
    \cline{2-5}\cline{7-10}\cline{12-15}\cline{17-20}
    &SR$\uparrow$ &AT$\downarrow$ &hN$\uparrow$ &AR$\uparrow$ & &SR$\uparrow$ &AT$\downarrow$ &hN$\uparrow$ &AR$\uparrow$ & &SR$\uparrow$ &AT$\downarrow$ &hN$\uparrow$ &AR$\uparrow$& &SR$\uparrow$ &AT$\downarrow$ &hN$\uparrow$ &AR$\uparrow$\\
    \midrule
    
    $\text{\;\;MaxE}^\dagger$   &0.061 &9.94 &0.015 &0.451
    &  & 0.067 & 9.91  & 0.017 &0.199  && 0.846  & 6.52  & 0.278 &0.953 && 0.015 & 9.98 & 0.004 & 0.324 \\
    
    $\text{\;\;EAR}^\dagger$& 0.117 & 9.78 & 0.037 &0.254  && 0.059  & 9.94  & 0.015 & 0.197  && 0.804  & 6.91  &0.273 &0.956 && 0.054 &9.95  &0.009 &0.353\\
    
\midrule[0.08pt]
    $\text{\;\;SCPR}^{\dagger-}$& 0.112 & 9.77 & 0.036 & 0.262  && 0.072  & 9.88  & 0.023 & 0.208  && 0.782 & 6.89  &0.272 &0.977 && 0.058 & 9.94 & 0.015 & 0.362\\

    $\text{\;\;SCPR}^\dagger$& 0.128 & 9.69 & 0.042 & 0.273  && 0.075  & 9.82  & 0.027 & 0.248 && 0.826 & 6.85  &0.301 &\underline{0.985} && 0.091 & 9.84 & 0.023 & 0.371\\    
\midrule[0.08pt]
    
    $\text{\;\;FPAN}^{\dagger-}$ &0.167 &9.74 &0.039 &0.266 &&0.067  &9.90  &0.022 &0.214  &&0.880  &5.94  &0.297  &0.966 && 0.071 &9.91 &0.017 & 0.354\\
    
    $\text{\;\;FPAN}^\dagger$ &0.264 &9.54 &0.052 &0.313 && 0.069  &9.89  &0.028 & \underline{0.365} &&\underline{0.895}  &\underline{5.87}  &0.304  &0.972 &&0.082 &9.88 &0.022 & 0.361\\
\midrule[0.08pt]

    $\text{\;\;UNI}^{\dagger-}$  &0.243 &9.58&0.050 &0.413 &&0.158 &9.64 &0.042 &0.129 &&0.805 &6.57 &0.289 &0.950 && 0.124 &9.80 &0.032 &0.319\\

    $\text{\;\;UNI}^\dagger$  &\underline{0.286} &\underline{9.48} &\underline{0.068} &\underline{0.475} &&\underline{0.196} &\underline{9.45} &\underline{0.062} &0.178
    &&0.830 &6.42 &\underline{0.312} &0.954 && \underline{0.162} &\underline{9.76} &\underline{0.052} &\underline{0.382}\\
    \midrule
    \;\;Ours &{\bfseries 0.547}& {\bfseries 8.88} &{\bfseries 0.170 }&{\bfseries 0.855 }  &&{\bfseries 0.291} &{9.47}&{\bfseries 0.089}&{\bfseries 0.439} &&{\bfseries 0.924} &{\bfseries 5.81} &{\bfseries 0.349}& { 0.968} &&{\bfseries 0.277} & {\bfseries 9.75} & {\bfseries 0.081} &{\bfseries 0.778}\\

    \bottomrule
    \end{tabular}}
    \vspace{-0.2cm}
\end{table*}

\vspace{-0.15cm}
\subsection{Experimental Setup}
\vspace{-0.1cm}
\subsubsection{Dataset.} 
To evaluate the proposed method, we adopt four existing MCR benchmark datasets. The details of these datasets are given in Table \ref{tab:dataset}.
\begin{itemize}[leftmargin=*]
    \item \textbf{LastFM} and \textbf{LastFM*} \cite{SCPR}. The LastFM dataset is used in assessing music artist recommendations. To facilitate modeling, original features are merged into $33$ coarse-grained groups.
    Considering the unfeasibility of utilizing knowledge graphs to manually merge attributes in real situation, the initial features are used in the LastFM* dataset.
    \item \textbf{Yelp} and \textbf{Yelp*} \cite{SCPR}. The Yelp dataset is for the business recommendation field, involving two-tier taxonomy with $29$ first-tier categories \cite{EAR}.
While Yelp* is a fine-grained version with $590$ second-tier category attributes.
\end{itemize}

\vspace{-0.2cm}
\subsubsection{User Simulator.}\label{sec:usersim}
Due to the difficulty of interacting with real users, we design a user simulator to train and evaluate MCR frameworks. 
Following previous works \cite{EAR,SCPR,Unicorn}, the classical user simulator is adapted to our MSMCR scenario from the construction of a session and the generation of activation attributes. 
Note that this paper mainly focuses on the scenario itself and its solution, and a more reasonable user simulator can be designed for MSMCR.

\begin{itemize}[leftmargin=*]
    \item \emph{The Construction of a Session.}  We simulate a session composed of ordered subsessions from a chronological user-item interaction sequence.
    $i$-th subsession $s^{i}_u$ is constructed by $i$-th interacted item $v^i$ of user $u$ correspondingly.
    For each subsession $s^i_u$ in session $S_u$, we regard the item $v^i$ as the ground-truth target item, and its attribute set $\mathcal{A}_{v^i}$ as the oracle attributes preferred by the user. 
    The session length $n$ is defined as $min(Random(N_{min},N_{max}),M_{u})$, where $M_u$ is the interaction length of user $u$, $N_{min}$ and $N_{max}$ are the threshold values of session length $n$ to align with realistic conversation session size.
    We report the mean value of $n$ resulted from our simulation in Table \ref{tab:dataset}.  
    \item \emph{The Generation of Activation Attributes.} The activation attributes $\mathcal{A}_{v^n}^*$ of the current subsession within a session are the subset of oracle attributes $\mathcal{A}_{v^n}$. 
    Specifically, it is defined as the top-two-ranked attributes based on user-attribute affinity score: $\mathbf{U}^\top_{u}\mathbf{A}_a + (\sum_{j=1}^{n-1}\mathbf{V}_{v^j}^\top\mathbf{A}_a) / (n - 1), \forall  a \in \mathcal{A}_{v^n}$ for each session, following the design of related works \cite{MetaCRS,TGCN}.
    In this formula,  $v^j$ represents the target item of the $j$-th subsession in session $S_u$, and $\mathbf{U}$,$\mathbf{V}$,$\mathbf{A}$ are pre-trained embeddings of users, items and attributes, respectively, which are obtained via TransE \cite{TransE} in \cite{Unicorn}. 
    Note that these activation attributes are applied uniformly across all methods, and their validity will be confirmed in Section \ref{sec:human_eval}.
\end{itemize}

\vspace{-0.15cm}
\subsubsection{Baseline Models.} \label{sec:baseline}
To verify model performance, our proposed model is compared with the following five representative multi-round CRS methods.
\begin{itemize}[leftmargin=*]
    \item \textbf{Max Entropy} (\textbf{MaxE}) \cite{EAR} chooses to ask the attribute with maximum entropy among the candidate items or recommend the top-ranked items with a certain probability.
    \item \textbf{EAR} \cite{EAR} proposes an Estimation-Action-Reflection three-stage solution, which achieves more accurate results than traditional algorithms like Abs Greedy \cite{Abs_Greedy} and CRM \cite{CRM}.
    \item \textbf{SCPR} \cite{SCPR} is designed to use interactive path reasoning on the graph to trim candidate attributes and adopt the DQN \cite{DQN} framework for action selecting.
    \item \textbf{FPAN} \cite{FPAN} extends EAR by adapting item and user embedding via the gate mechanism based on online user feedback.
    \item \textbf{UNICORN} (\textbf{UNI}) \cite{Unicorn} constructs a unified policy learning framework based on a dynamic weighted graph.
\end{itemize}

As discussed in Section \ref{sec:relate_works}, several recent works (e.g., MIMCR \cite{MCMIPL}, Bundle-MCR \cite{BundleMCR}, GPR \cite{GPR}, and MetaCRS \cite{MetaCRS}) primarily focus on the different novel MCR scenarios, which go beyond the scope of our MSMCR setting. Therefore, we do not adopt them.
Also, we do not use the recent work CRIF \cite{CRIF}, since main contributions of CRIF are based on Inverse Reinforcement Learning  \cite{irl} for conversation policy learning, which differs from the general fixed reward setting in other baselines and our MSCAA (cf. Section \ref{sec:details}).

For a fair comparison, we revise all the above baselines as follows: 
(1) We utilize previous subsessions information to complement the user presentation for the item and attribute scoring. 
Specifically, we first apply GRU to obtain the representation of user item-level interest from previous subsessions, aligned with our method (cf. Section \ref{sec:interest-model}).
Then, the original user representation and item-level interest presentation are further fused as the updated user representation via the gating mechanism \cite{gating,pang}.
(2) We employ our user simulator for all baselines.

Note that all baseline implementations are modified to adapt to the MSMCR scenario, and therefore these methods are named with superscript $\dagger$. 
In addition, to further verify the necessity of previous subsessions, we remove the implementation of (1) and only preserve the setting of activation attributes (2) for several baselines which are named with $\dagger-$.

\vspace{-0.15cm}
\subsubsection{Evaluation Metrics.}
Following previous studies on MCR \cite{EAR,SCPR}, the accumulative ratio of successful conversational recommendation assessed by SR@$T$ (\emph{success rate at turn $T$}) and the average number of turns for all subsessions assessed by AT (\emph{average turn}) are adopted. 
Similar to SR@$T$, we introduce AR@$T$ (\emph{activation rate at turn $T$}) to measure the accumulative ratio of hitting activation attributes in the session.
Moreover, we also adopt hN@($T$,$K$) to measure the recommendation accuracy following \cite{Unicorn}, which considers the rank performance in both list- and turn-level.
We omit the $@T$ and $K$ terms in the following experiments for simplicity.
The higher value of SR, AR, and hN indicates better performance, while a lower AT indicates higher effectiveness.
Notably, SR, AT and our proposed AR only consider a certain aspect of CRS \cite{Unicorn}, while hN is a comprehensive metric for evaluating the overall framework.

\vspace{-0.15cm}
\subsubsection{Implementation Details.}\label{sec:details}

We implement the proposed method based on PyTorch. 
We randomly divide interactions across each dataset into training, validation, and test parts with the ratio of $7:1.5:1.5$ and then separately generate the session samples based on by user simulator described in Section \ref{sec:usersim}. 
The maximum/minimum session length $N_{max}$, $N_{min}$ is set to $4$, $2$ respectively.
We set the maximum turn $T$ and the size of the recommendation list $K$ as $10$. 
The embedding dimension $D$ is set as $64$.
We employ the Adam/SGD optimizer to train the recommendation/policy module with the learning rate $5e^{-4}$ and $1e^{-3}$ separately.  
The heterogeneous graph $G$ is constructed by the training dataset.
The number of RGCN layers $L_g$ and Transformer layers $L_t$ are all set to be $2$. 
Discount factor $\gamma$ is set to be $0.7$. 
The temperature parameter $\tau$ is set with $0.5$. The weight $\omega$ of InfoNCE constraint is set to be $0.01$.
We set six rewards 
$[r_{rec\_acc}, r_{rec\_rej},r_{ask\_acc},r_{ask\_rej},r_{ask\_unk},r_{quit}]=[1, -0.1, 0.01, -0.1, -0.1, -0.3]$ following reward settings in \cite{SCPR}.

\begin{figure}[t]
    \centering
    \begin{subfigure}{0.48\linewidth}
        \includegraphics[width=\textwidth]{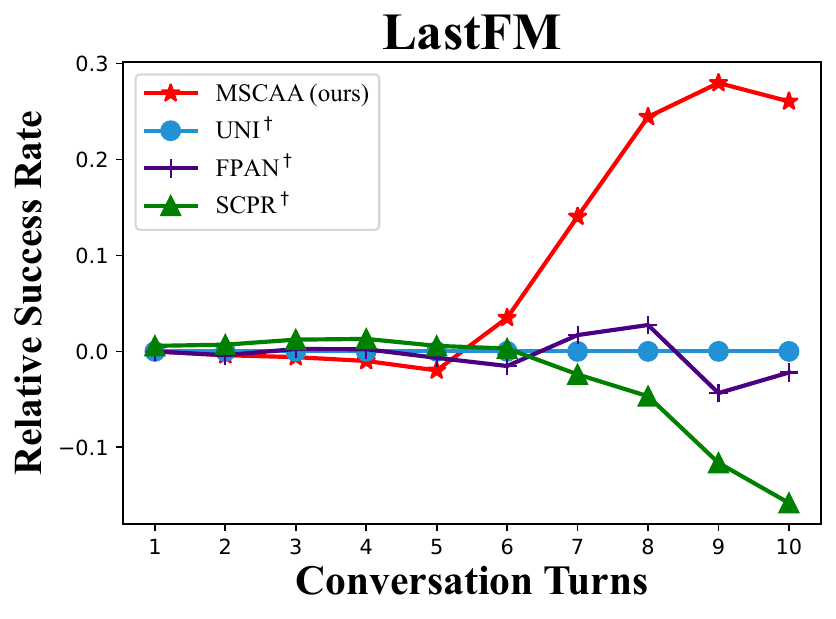}
    \end{subfigure}
    \begin{subfigure}{0.48\linewidth}
        \includegraphics[width=\textwidth]{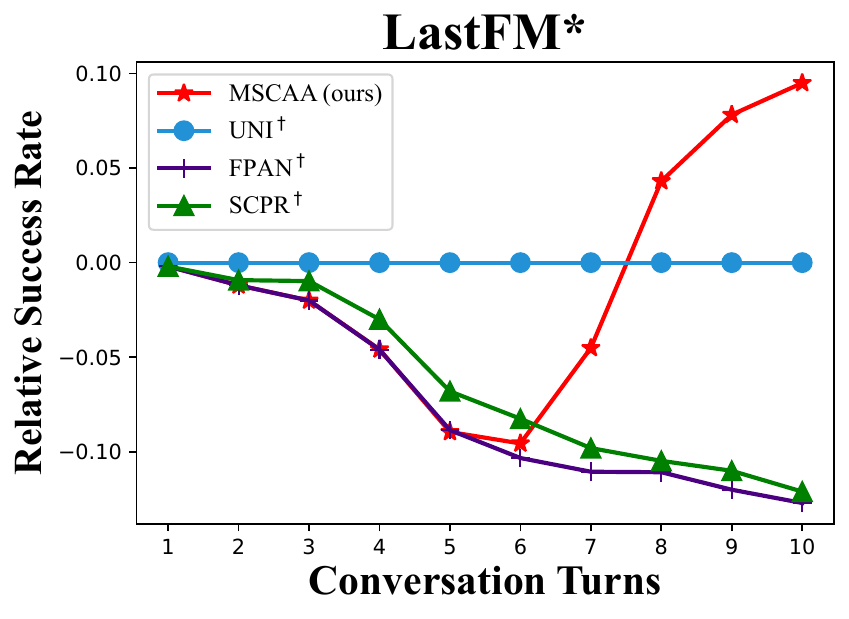}
    \end{subfigure}
    \begin{subfigure}{0.48\linewidth}
        \includegraphics[width=\textwidth]{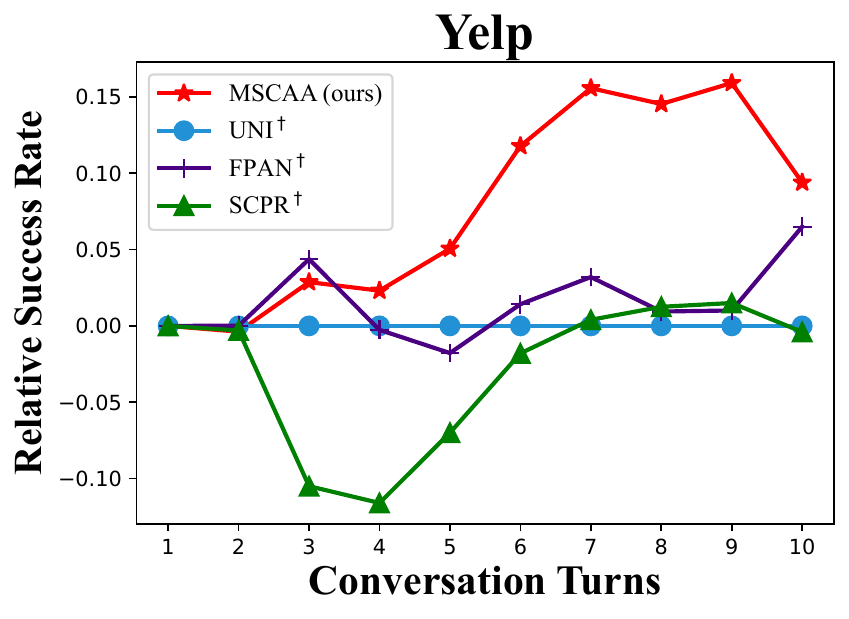}
    \end{subfigure}
    \vspace{-0.2cm}
    \begin{subfigure}{0.48\linewidth}
        \includegraphics[width=\textwidth]{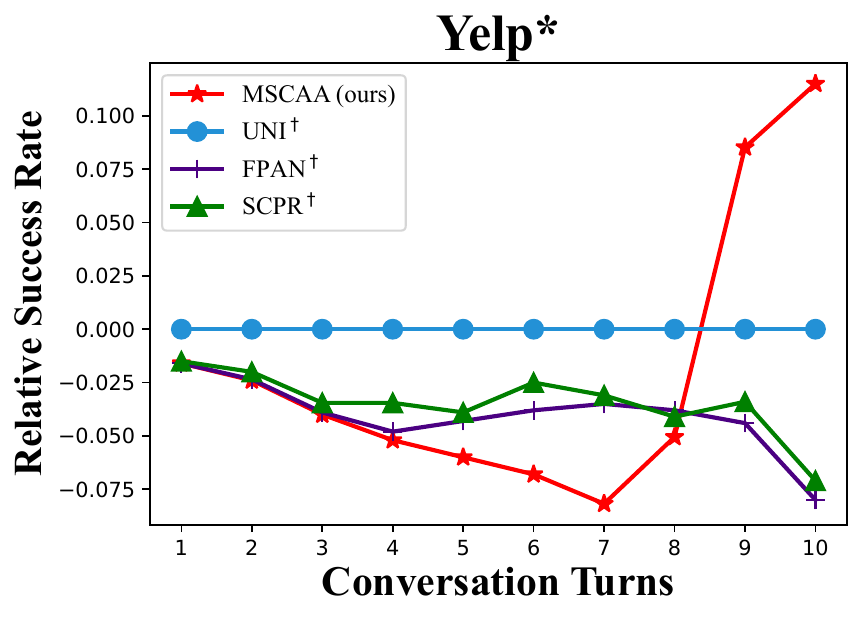}
    \end{subfigure}
     \vspace{-0.2cm}
    \caption{Comparisons of Relative Success Rate at Different Conversation Turns.}
    \label{fig:overall}
    \vspace{-0.5cm}
\end{figure}

\vspace{-0.2cm}
\subsection{Performance Comparison}\label{sec:exper}
\vspace{-0.05cm}
\subsubsection{Overall Comparison}
To demonstrate the overall performance of the proposed model, we compare it with the adapted SOTA methods for the MSMCR scenario. 
Table \ref{tab:overall} shows the comparison of experimental results between the baseline models and our proposed method. 
We can obtain the following notable observations:

\begin{itemize}[leftmargin=*]
    \item In most cases, our proposed method exhibits significant advantages among all baselines on four datasets. The main reason behind this is that MSCAA carefully combines three dedicated modules, especially the recommendation module that comprehensively models different user interests from heterogeneous information sources, to hit the user desired item in a few turns of subsession.
    \item Interestingly, several baselines (e.g., $\text{SCPR}^\dagger$ and $\text{UNI}^\dagger$) sometimes outperform MSCAA in AR and AT metrics. 
    It could be attributed to the policy bias in these models that prefer to ask or recommend. 
    For instance, more asking times will result in better accumulative performance on AR.
    However, our method still outperforms them in the comprehensive metric (i.e., hN) with a scalability trade-off between asking and recommendation.
    \item Our method's improvement is relatively limited on the Yelp dataset among all four datasets. 
    In addition, the performance of both LastFM* and Yelp* is comparatively inferior to the other two datasets. 
    These could be explained that the setting of the Yelp dataset is to ask enumerated questions \cite{EAR}, which makes it easier to hit the activation attributes and then sharply diminish the candidate item space for all methods.
    However, for the LastFM* and Yelp* datasets, it is harder to ask appropriate attributes to activate user interests due to the numerous candidate attributes.
    \item Compared to ``${\dagger-}$'', ``$\dagger$'' version method achieves a better overall performance, which indicates that it is effective to utilize the information of previous subsessions via sequential modeling in the MSMCR scenario.
\end{itemize}

\subsubsection{Comparison at Different Conversation Turns}
Besides SR@$\emph{10}$, we also present a detailed performance comparison of success rate at each turn in Figure \ref{fig:overall}.
To facilitate the observation, we only report the relative success rate of four representative methods compared with the SOTA baseline UNI$^\dagger$ in our MSMCR setting. 
For example, in Figure \ref{fig:overall}, the blue line of UNI$^\dagger$ is set to $y = 0$.
As can be seen, UNI$^\dagger$ or even other baselines may outperform MSCAA in the first few turns, while our MSCAA achieves a pretty performance at the end stage of the conversation. 
This phenomenon can be explained that MSCAA is more inclined to ask appropriate attributes to elicit user interest at first, and after that, can more confidently recommend items to satisfy the user's current demand. 
Conversely, the other baselines may recommend items successfully instead of asking more attributes in the earlier turns since they do not consider the activation of user interest. 
However, due to the lack of acquisition of the current user interest, they are inferior to MSCAA lastly.

\vspace{-0.15cm}
\subsection{Ablation Study}\label{sec:ablation}
\vspace{-0.05cm}
To verify the design effectiveness of MSCAA, we remove or replace key modules one-by-one and report results in Table \ref{tab:abl}.

\noindent \textbf{Impact of recommendation modules.}
The results in (a-c) show that the missing of any type of information (i.e.,
historical interactions, previous subsessions, or the online user feedback information) causes the degradation of model performance.
Especially, the long-term interest is the most significant among these three for highlighting the personality of users.
Moreover, (d) demonstrates the significance of contrastive loss for the recommendation.
(e) suggests that unified graph modeling is pivotal for long-term interest extraction.
(f) represents the GRU modules in Section \ref{sec:interest-model} are replaced with ``mean'' operation, which indicates the importance of sequential encoder of previous subsessions for recommendation in our MSMCR scenario.
Based on the above observations, we manifest that any input information and main components are indeed essential in the context-aware recommendation module.

\noindent \textbf{Impact of Attribute and Conversation Policy.}
(g) and (h) indicate that our attribute selection policy is replaced by Max Entropy and Max User-like score rule, respectively. 
The results show the effectiveness of our attribute agent for selecting an appropriate attribute based on the conversation state.
And (i) means the fine-grained score information $\boldsymbol{s}_{item} \oplus \boldsymbol{s}_{attr}$ in conversation state is masked. From the result, this kind of information is necessary for a high-level action decision.

\begin{table}
\setlength{\abovecaptionskip}{0cm}
\setlength{\belowcaptionskip}{-0.1cm}
\caption{Results of the Ablation Study. }
\label{tab:abl}
\vspace{0.1cm}
\begin{subtable}[]{0.48\textwidth}
\centering
 \resizebox{0.99\textwidth}{23mm}{
  \begin{tabular}{p{3.5cm}<{\raggedright}p{0.6cm}<{\centering}p{0.6cm}<{\centering}p{0.6cm}<{\centering}p{0.01cm}<{\centering}p{0.6cm}<{\centering}p{0.6cm}<{\centering}p{0.6cm}}
    \toprule
    \multirow{2}{*}{\bfseries{\;\;\;\;\;\;\;\; Models}}& \multicolumn{3}{c}{\textbf{LastFM}}&& \multicolumn{3}{c}{\textbf{Yelp*}}\\
    \cline{2-4} \cline{6-8}
    &SR$\uparrow$ &hN$\uparrow$ &AR$\uparrow$ &&SR$\uparrow$ &hN$\uparrow$ &AR$\uparrow$\\ 
    \midrule
    \textbf{(a)} w/o Long-term Interest &0.255& 0.078&0.490  &&  0.065& 0.019&0.450\\
    \textbf{(b)} w/o Short-term Interest &0.463&0.143& 0.844 && 0.146  & 0.043&0.550\\
    \textbf{(c)} w/o Current Interest &0.537& 0.167&0.845 &&  0.178& 0.052&0.720\\
    \textbf{(d)} w/o InfoNCE Loss &0.538& 0.168&0.852  &&  0.265 &  0.078 &0.759\\
    \textbf{(e)} w/o RGCN &0.517 &0.161 & 0.838 && 0.261  &0.076 & 0.760\\
     \textbf{(f)} w/ Mean &0.537&0.166&0.853  && 0.262  & 0.077& 0.764 \\
     \midrule
    \textbf{(g)} w/ Max Entropy &0.102 &0.030&0.333 &&0.032  &0.009&0.438 \\
    \textbf{(h)} w/ Max User-like &0.422& 0.132& 0.851 &&  0.138& 0.041&0.775\\
    \textbf{(i)} w/o $\boldsymbol{s}_{item} \oplus \boldsymbol{s}_{attr}$ & 0.538 & 0.169&0.848  && 0.270 & 0.079&0.770 \\
    \midrule
    \textbf{MSCAA} &\bfseries 0.547 & \bfseries 0.170 & \bfseries 0.855 && \bfseries 0.277& \bfseries 0.081 & \bfseries 0.778 \\
    \bottomrule
    \end{tabular}}
\end{subtable}
\vspace{-0.2cm}
\end{table}

\vspace{-0.15cm}
\subsection{Case Study}\label{sec:case}
\begin{figure}[t]
    \centering
    \begin{subfigure}{1.0\linewidth}
        \includegraphics[width=\textwidth]{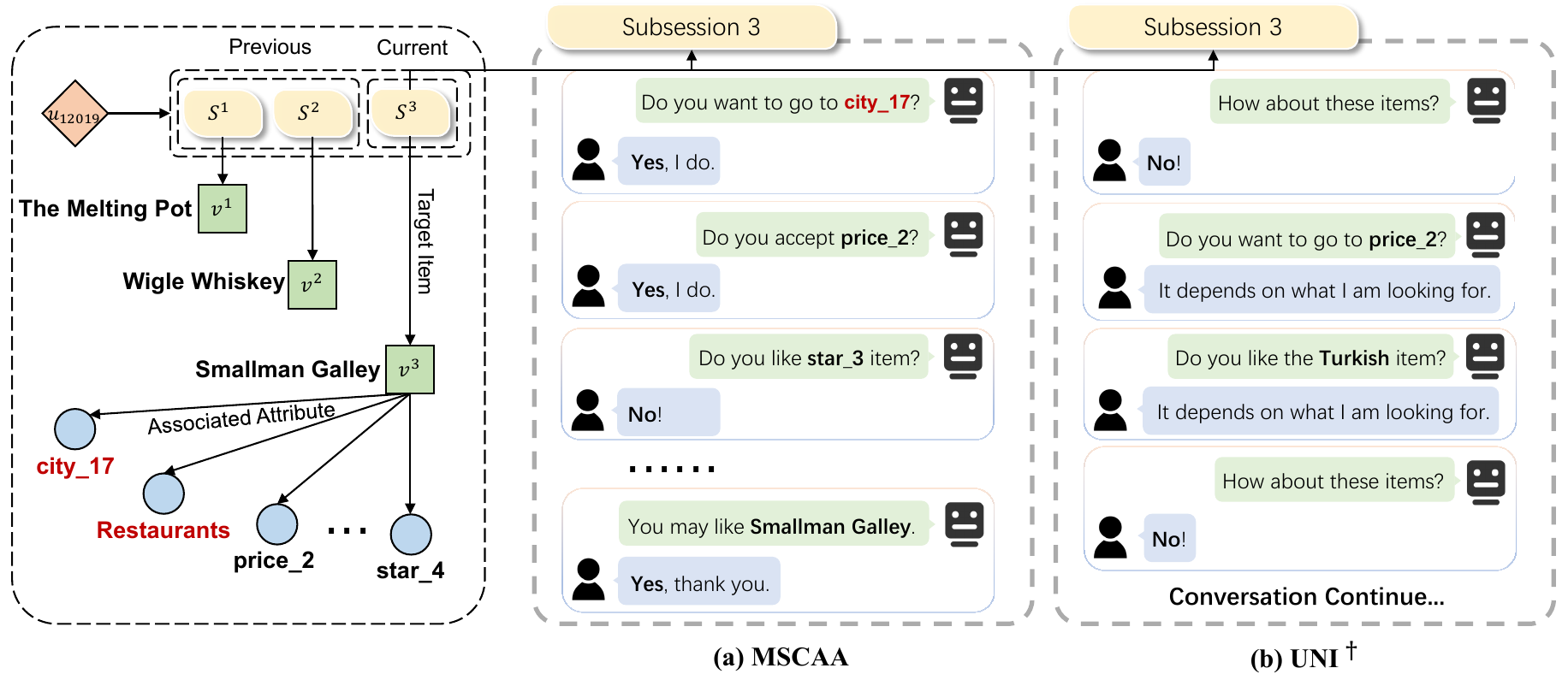}
    \end{subfigure}
    \vspace{-0.3cm}
    \caption{A session case generated by our framework MSCAA and UNI$^\dagger$. 
    The left side illustrates the sampled user and his/her interacted three subsessions with corresponding target items and associated attributes, in which activation attributes are red bold fonts.
    We present the conversation process of the current subsession (i.e., subsession $3$) in detail.
    }
    \label{fig:case_study}
    \vspace{-0.2cm}
\end{figure}

\vspace{-0.05cm}
To show the process of activating user interest from the system side and then recommending successfully, we further present a session case generated by our framework MSCAA and UNI$^\dagger$ based on the same test instance from Yelp* dataset in Figure \ref{fig:case_study}.
This session consists of three subsessions where in the previous two subsessions, the user $u_{12019}$ has already obtained his/her desired items \emph{The Melting Pot} and \emph{Wigle Whiskey} which are related to food and drink in \emph{city\_17}, and he/she continues the third subsession to seek more advice.
For the current subsession $3$, two main attributes \emph{city\_17} and \emph{Restaurants} of the target item \emph{Smallman Galley} are considered as the activation attributes of this dialogue.

It can be observed that based on the comprehensive information especially the previous subsessions, our MSCAA precisely hits the activation attribute \emph{city\_17} to elicit user interest for subsequent specific attribute-asking questions.
As the subsession turn progresses, the user demand is gradually crystallized for assisting the system in seeking a satisfactory item (also a restaurant) successfully.
While the baseline UNI$^\dagger$  recommends items at the beginning of the subsession 
since it utilizes user interest from previous subsessions to be confident to choose this action.
However, it fails because of the missing of  current user interest in this subsession.
And then, UNI$^\dagger$ can present an attribute \emph{price\_2} that satisfies the user to a certain extent, but this over-specific question may confuse the user before his/her current interest is activated.
In summary, compared to UNI$^\dagger$, MSCAA can not only ask appropriate attributes to elicit user interest quickly in the early turns but make a comprehensive decision of recommendation or asking, 
beneficial for succeeding.

\vspace{-0.15cm}
\subsection{Human Evaluation}\label{sec:human_eval}
\vspace{-0.05cm}
Although the above experiments demonstrate the effectiveness of our framework in the proposed user settings, we further conduct a human evaluation to answer two questions:
(1) Are the activation attributes appropriate in our user simulator? 
(2) Does our method outperform other baselines in the realistic scenario?
Specifically, we randomly select $100$ session samples from Yelp* from our framework and the competitive baseline $\text{UNI}^\dagger$.
The experiment involves $20$ post-graduate volunteers who evaluate $10$ samples each, with each sample being evaluated by $2$ different volunteers.
For the first question, volunteers need to review the user's previous subsessions and choose $2$ main attributes of the target item in the current subsession as human-labeled activation attributes.
We then calculates Jaccard Similarity \cite{levandowsky1971distance} for each sample between human-labeled and our simulated activation attribute set, and the average value is $0.73$.
For the second question, the volunteers are required to browse the conversation process of the current subsession and select the better conversation from $<\text{MSCAA},\text{UNI}^\dagger>$ episode pairs in their subjective view, following the human evaluation method in \cite{BundleMCR}. 
Lastly, we collect $200$ results: $<\text{MSCAA},\text{UNI}^\dagger>$ votes are $127:73$. 
Overall, these results validate the quality of our user simulator and the effectiveness of our method in the realistic scenario.

\vspace{-0.15cm}
\section{Conclusion}
\vspace{-0.1cm}
In this paper, we extend the MCR to a general CRS setting, MSMCR 
in which the user continues the dialogue with the system after several subsessions and might preserve no clear interest in the current subsession,
and the system would proactively take actions to activate the user's dynamic interest.
For this scenario, a novel framework called MSCAA is introduced to model user interests comprehensively for the recommendation, learn the flexible strategy for asking the appropriate attributes, and manage actions between asking and recommending adaptively.
Extensive experimental results on four adapted datasets verify the effectiveness and superiority of our framework in the proposed scenario.

\bibliographystyle{ACM-Reference-Format}
\bibliography{ref}
\appendix

\section*{Ethical Considerations}
It is believed that our proposed MSMCR scenario and the corresponding solution framework MSCAA could mitigate some challenging ethical problems in several conventional recommender systems (e.g., collaborative filtering, sequential recommendation, etc.).
The MSMCR scenario, which inherits but surpasses the existing MCR setting, has the following merits:
\begin{itemize}[leftmargin=*]
\item MCR emphasizes the dynamic and precise acquisition of user preferences during \textit{online} conversations, which is ideally less reliant on learning from offline interactions that might be noisy and bring about unintentional biases (e.g., popularity bias).
Furthermore, our MSMCR additionally considers the \textit{previous subsessions} of one session for users, comprehensively excavating online information except for feedback in the current subsession of the session.
As such, MSMCR could further mitigate the bias of offline interaction data.
\item Our MSMCR also takes more realistic characteristics of users into consideration. Specifically, users might continue to engage in the conversation session aimlessly until a raised topic triggers user interests. Therefore, the \textit{activation attributes} of users are designed in the subsequent subsession of one session, so as to resemble real-world situations for satisfying user demands.
\end{itemize}

As for our solution, MSCAA framework is established on reinforcement learning, therein the \textit{attribute selection} and \textit{conversation policy} is learned by policy learning.
Different from traditional predictive and matching methods, policy learning aims to maximize cumulative gains for all conversations, enabling to cater to long-term user satisfaction (i.e., the successful conversation).
This goal is more user-friendly to enhance the overall user experience.

However, due to the difficulty of interacting with real users, collecting online user feedback is too ideal to train MSCAA for alleviating the unintentional biases in the offline interaction data.
Hence, following the previous related works, we still adopt a adapted user simulator based on user interaction data, including the construction of one session, the generation of activation attribtues and the simulation of user feedback.
Indeed, it is undeniable that using such a simulated environment introduces potential discrepancies and biases compared to the real-world distribution.
Significant further progress needs to be made in areas like Sim2Real (Simulation to Reality) before we deploy this system in a production setting.

\clearpage

\section{Main Notations}

\begin{table}[htbp]
    \setlength{\abovecaptionskip}{0cm}
    \setlength{\belowcaptionskip}{-0.1cm}
    \caption{Main notations in the paper.}
    \label{tab:notations}
    \vspace{0.1cm}
    \centering
    \begin{tabular}{|c|m{0.7\linewidth}|}
    \hline
    \textbf{Notations} & \textbf{Definitions and Descriptions} \\
    \hline
    $u$ & A user \\
    \hline
    $\mathcal{U}$ & The set of all users\\
    \hline
    $v$ & An item \\
    \hline
    $\mathcal{V}$ & The set  of all items\\
    \hline
    $a$ & An attribute \\
    \hline
    $\mathcal{A}$ & The set of all attributes \\
    \hline
    $\mathcal{A}_v$ & The set of attributes that belong to an item $v$ \\
    \hline
    $s$ & A subsession \\
    \hline
    $S_u$ & A session interacted by user $u$ \\
    \hline
    $P_{S_u}$ & The previous subsessions of session $S_u$ \\
    \hline
    $P^{\mathcal{V}}_{S_u}$ & The sequence of the target items of previous subessions in the session $S_u$ \\
    \hline
    $P^{\mathcal{A}}_{S_u}$  & The sequence of the target attribute sets of previous subsessions in the session $S_u$\\
    \hline
    $s_u^n$ & The current (i.e., $n$-th) subsession of session $S_u$, where $n$ is the length of the session \\
    \hline
    $v^n$ & The target item of subsession $s^n_u$ (also session $S_u$) \\
    \hline
    $\mathcal{A}_{v^n}^*$ & The activation attributes of subsession $s^n_u$ (also session $S_u$)\\
    \hline
    $\mathcal{A}_{acc}$ & The set of attributes accepted by a user in the current subsession\\
    \hline
    $\mathcal{A}_{rej}$ & The set of attributes rejected by a user in the current subsession \\
    \hline
    $\mathcal{A}_{cand}$ & The set of candidate attributes in the current subsession \\
    \hline
    $\mathcal{V}_{cand}$ & The set of candidate items in the current subsession \\
    \hline
    $\mathcal{V}_{rej}$ & The set of items rejected by a user in the current subsession \\
    \hline
    $\boldsymbol{s}_{\mathfrak a}$ & The state of attribute selection policy\\
    \hline
    $\boldsymbol{s}_{\mathfrak c}$  & The state of conversation policy\\
    \hline
    $\pi_{\mathfrak c}$  & The conversation policy\\
    \hline
    $\pi_{\mathfrak a}$  & The attribute selection policy\\
    \hline
\end{tabular}
\end{table}

\section{Further Discussion}

\subsection{Differences between MSMCR and Existing Scenarios}

Our MSMCR scenario aims to activate user interests in the current subsession for the successful recommendation by considering the correlation between previous subsessions within a session.

We highlight that it differs from two related recommendation fields: (1) sequential/session recommendation \cite{IAGNN,pang}, which lacks interactive dialogue between users and the system and only provides item recommendation once; (2) MCR \cite{EAR,SCPR,Unicorn}, which disregards the previous subsessions and activation attributes.

\subsection{The Strength and Weakness on the Simulator}

Regarding the strengths, it is commendable that our proposed user simulator can handle more general behavior and preference of users (e.g., the recent subsessions imply short-term user interest), meanwhile providing more realistic feedback (e.g., ``unknown'' when vague interest) than previous CRS simulators \cite{EAR,SCPR,Unicorn}.

As for the weaknesses, it is understandable that the user simulator's reliance on activation attribute extraction may lead to some deviation in the distribution of simulated user feedback from reality. 
Specifically, 
(1) some attributes that users really like may not have been generated, resulting in the user simulator lacking more explicit responses;
(2) some noisy attributes may have been mixed in with certain users' preference attributes, causing the user simulator to conduct wrong dialogues for activating current user interest.
Additionally, the construction of a session is based on the chronological user-item interaction sequence, not consecutive online dialogues between users and the system, which may also bring deviations that are somewhat inconsistent with reality.
As for the setting of hyper-parameters on conversations (cf. Section \ref{sec:details}), although we follow the existing CRS works \cite{MetaCRS,FacTCRS} to set the maximum turn $T$ as a fixed value $10$, trying to be as realistic as possible \cite{yang2017neural}, this is too ideal to introduce deviations inevitably.
The same is for the maximum/minimum session length.

\end{document}